\documentclass[12pt]{article}
\usepackage{epsfig}

\parskip=\medskipamount 
plus.3em minus.5em \abovedisplayshortskip=.5em plus.2em minus.4em
\renewcommand{\thesection}{\arabic{section}}

\catcode`@=11

\@addtoreset{equation}{section} \@addtoreset{equation}{subsection}
\def\theequation{\ifnum\value{section}=0 \arabic{equation}\ignorespaces
\else \ifnum\value{section}=-1 A.\arabic{equation}\ignorespaces
\else \ifnum\value{subsection}=0
\thesection.\arabic{equation}\ignorespaces \else
\thesection.\arabic{subsection}.\arabic{equation}\ignorespaces
                             \fi
                        \fi
                   \fi}

{\catcode`\'=\active \def'{{}^\bgroup\prim@s}}

\catcode`@=12

\newcommand{\bq}{\begin{equation}}
\newcommand{\be}{\begin{equation}}
\newcommand{\fq}{\end{equation}}
\newcommand{\ee}{\end{equation}}
\newcommand{\bqr}{\begin{eqnarray}}
\newcommand{\beqs}{\begin{eqnarray}}
\newcommand{\fqr}{\end{eqnarray}}
\newcommand{\eeqs}{\end{eqnarray}}

\newcommand{\rf}[1]{(\ref{#1})}

\def\bop#1{\setbox0=\hbox{$#1M$}\mkern1.5mu
    \vbox{\hrule height0pt depth.04\ht0
    \hbox{\vrule width.04\ht0 height.9\ht0 \kern.9\ht0
    \vrule width.04\ht0}\hrule height.04\ht0}\mkern1.5mu}

\begin{document}
\thispagestyle{empty}

\begin{flushright}
\begin{tabular}{l}
hep-th/0508037 \\
\end{tabular}
\end{flushright}

\vskip .6in
\begin{center}

{\bf Perturbative Solution to D-terms in N=2 Models and Metrics}

\vskip .6in

{\bf Gordon Chalmers}
\\[5mm]

{e-mail: gordon@quartz.shango.com}

\vskip .5in minus .2in

{\bf Abstract}

\end{center}

$N=2$ gauged non-linear sigma models are examined classically and their 
D-terms are solved.  The variation of the classical Lagrangian in order 
to solve for the auxiliary fields is identical to integrating these modes 
functionally.  The latter is performed  
for the general quotient.  The D-term solution is equivalent to solving, 
a coupled set of algebraic equations.  

\vfill\break

\vskip .2in
\noindent {\it Introduction}

The solution to the D-terms in ${\cal N}=2$ non-linear sigma models is 
useful in the construction of metrics on toric varieties as well as in 
the construction of particle physics and string models.  Their solution is 
problematic because a very general set of algebraic equations has to be 
solved.  

The K\"ahler metrics are also required in order to generate geodesic 
flows on these spaces.  The solutions to these flow equations are 
required in order to find solutions to algebraic systems \cite{Chalmers1} 
and systems of non-linear differential equations \cite{Chalmers2}.

The solution to these D-term equations can be achieved by a counting of tree 
diagrams in scalar field theory models \cite{Chalmers3}.  The scalar field 
model contains 
interactions of an arbitrary degree in the external lines, due to the 
exponential interactions in the auxiliary fields.  The counting can be 
performed in principle by a careful examination of the auxiliary field 
interactions and is torically model specific.  However, a deformation can 
be added to the Lagrangian which simplifies this count 
quite much; after the solution to the D-terms, the deformation parameters 
are taken to zero.  (The same tree graphs are used in the full quantization 
of scalar field and gauge theories \cite{Chalmers4}-\cite{Chalmers8}. 

The general auxiliary components of the ${\cal N}=2$ Lagrangian is described 
by (see for example \cite{Kimura} also for several completed examples), 

\bqr 
{\cal L} = \sum_{i=1}^a \vert\phi_i\vert^2 \prod_{j=1}^b e^{Q_i^j V_j} + 
 r_j V_j \ ,  
\label{dterms}
\fqr 
with an implicit sum on the auxiliary components $V_j$ in the exponential.  
The field equations that are derived from \rf{dterms} are the set, 

\bqr 
r_j + \sum_{i=1}^a \vert\phi_i\vert^2 Q_i^j 
 \prod_{j=1}^b q_j^{Q_i^j}  = 0 \ .  
\fqr 
In some cases of these torics, field redefinitions may be used to find 
the K\"ahler potential; the general case and its algebraic system requires 
a different approach.

\vskip .2in 
\noindent{\it Solution} 

As opposed to solving a complicated system of algebraic equations, the 
Lagrangian and its integration over the auxiliary fields $V_j$ can be 
analyzed with the use of classical scattering diagrams which have no 
momentum structure.  There is no propagator or derivatives on the 
auxiliary fields, and as such, the integration of these fields is obtained 
by counting all of the inequivalent tree diagrams that contain external 
$(\phi_i,\bar\phi_i)$ lines and internal $V_j$ lines. 

The linear dependence on the $V_j$ field, as well as a non-trivial but 
required combinatoric property, is treated by changing its form as in, 

\bqr 
\sum_{j=1} r_j V_j \rightarrow \sum_{j,k} r_j r_k 
 \gamma_j \gamma_k (V_j+\alpha_j) (V_k+\alpha_k) \ .   
\fqr 
\bqr
= \sum r_j r_k \gamma_j \gamma_k V_j V_k + \gamma \sum r_j \gamma_j \alpha_j V_j 
+ \gamma^2 \qquad \gamma=\sum r_j \gamma_j \alpha_j  
\label{deformation}
\fqr 
A field redefinition of $V_j\rightarrow (V_j-\alpha_j)/r_j\gamma_j$ changes 
the deformation into 

\bqr 
\sum_{j=1} r_j V_j \rightarrow \sum_{i<j} V_i V_j \ .  
\fqr 
The limit of the linear term is obtained by taking $\gamma_j\alpha_j r_j=1$ 
with $r_j \gamma_j=0$, for example; another removal of the deformation can 
be obtained by field redefining $\phi_i\rightarrow \phi_i r_i \gamma_i$ 
followed by taking $\gamma_i$ to zero.  The latter is due to the counting of 
internal lines in the diagrams; due to the vertex structure there are always 
$n$ propagators with a diagram containing $n+2$ $\phi_i$ or $\bar\phi_i$ 
lines.  

This deformation in \rf{deformation} generates a uniformity to the 
combinatorics associated with the interacting auxiliary fields in 
the tree diagrams, as the 
propagator is completely symmetric in the indices $j$ and $k$.  As a result, 
the vertices are connected from one node to eachother in the most uniform 
manner possible; this effectively changes all of the $V_j$ into $V$ in the 
classical graphs so that only a single scalar field need be examined (while 
keeping the indices on the charges).  The removal of the deformation is 
obtained in the K\"ahler potential $K(\phi_i,\bar\phi_i)$, after its form 
is deduced from the classical scattering.

The complings in the Lagrangian \rf{dterms} are deduced by expanding the 
exponentials in the $V_j$ fields.  Each graph composed of the various $i$ 
vertices containing $p_i$ lines (two of which are external $\phi$ fields) 
are weighted with these couplings.  The K\"ahler potential is found from, 

\bqr 
K(\phi_i,\bar\phi_i)= \sum^{n_\phi/2} \sum_{\rm perms} 
 \prod \vert\phi_\sigma\vert^2 a_{n_\phi,m} \prod^{p_i} 
  \lambda_{\sigma,\{p_i\}} {1\over 2^{n_{\phi}/2} \prod p_i!} \ .  
\label{Kahler}
\fqr 
The counting number $a_{n,m}$ is used to count the number of diagrams 
containing the sum $\sum p_i = m$ and at $n_\phi$-point.  This counting 
function can be obtained by solving an associated function \cite{Chalmers1}.  
The potential is found in a patch around $\phi_i=\bar\phi_i=0$ and as an 
expansion which is not necessarily Taylor.   

The expansion of the exponentials in \rf{dterms} generates, 

\bqr 
\sum_{i=1}^a \vert\phi_i\vert^2 \prod_{j=1}^b e^{Q_i^j V_j} 
 = \sum_{i=1}^a \sum_{n=0}^\infty \vert\phi_i\vert^2 {1\over n!} 
  (\sum_{j=1}^b Q_i^j V_j)^n 
\fqr 
\bqr 
= \sum_{i=1}^a \sum_{n=0}^\infty \sum_{q_j=0} \vert\phi_i\vert^2 
  {1\over \prod q_j!} \prod_{j=1}^b (Q_i^j)^{n-q_j} (V_j)^{q_j} 
 \vert_{\sum q_j=n}  
\label{exponentexp}
\fqr 
The field redefinition of $V_i\rightarrow (V_j-\alpha_j)/r_j \gamma_j$ 
changes the form to, 

\bqr 
=\sum_{i=1}^a \sum_{n=0}^\infty \sum_{q_i=0}  \vert\phi_i\vert^2 
 \prod_{j=1}^b  
 \sum_{p_j=0}^{q_j} {1\over p_j! (q_j-p_j)!} (Q_i^j)^{n-q_j}
 (V_j)^{p_j} (-\alpha_j)^{q_j-p_j} {1!\over (\gamma_j r_j)^{q_j}}\ . 
\label{vertices}
\fqr 
The expansion \rf{vertices} leads to the vertex couplings, 

\bqr 
\lambda_{\sigma,\{p_i\}} = \sum_{q_i=0}  
 \prod_{j=1}^b {1\over (q_j-p_j)!} (Q_i^j)^{n-q_j} (-\alpha_j)^{q_j-p_j} 
{1!\over (\gamma_j r_j)^{q_j}} 
\label{vertex}
\fqr 
These couplings are used to weight the classical graphs, and 
find the K\"ahler potential.  The sum on $q_j$ extends at fixed $p_j$, 
with $\sum q_j = n$, with the sum on $n$.  These couplings together 
with the counting function $a_{2n_\phi,m}$ are used to find 
the K\"ahler potential with \rf{Kahler}; recall that $m=\sum (2+p_j)$, 
that is the count of lines at each vertex.  

After the computation, the deformation parameter is removed; without 
a field redefinition on the $\phi,\bar\phi$ coordinates, there could 
be a potential singularity and requires  
some summations to be performed; however the scaling of the coordinates 
removes this.  As $q_i$ becomes large, with the 
large order scaling $n\sim q_j$, the right hand side in \rf{vertex} 
tends to, 

\bqr 
\sum_{q_i=0} {1\over q_i!} (Q_i^j)^n  
 ({-\alpha_j\over Q_i^j \gamma_j r_j})^{q_j} (-\alpha_j)^{p_j} = 
 e^{Q_i^j} e^{-\alpha_j/Q_i^j \gamma_j r_j} (-\alpha_j)^{p_j} 
\fqr  
\bqr 
\sim e^{Q_i^j} (-\alpha_j)^{p_j} \ , 
\fqr 
This is a heuristic explanation of how the removal of deformation 
parameter can be explained by performing the sum in \rf{vertex}, 
and taking $\gamma_j r_j\rightarrow 0$.  The sum, and its asymptotic 
values has to be examined.  A field redefinition of the 
coordinates can handle the removable singularity in the $\alpha_j$.  
The above is one means to remove the deformation, but a direct 
scaling of the coordinates by the $r_i\gamma_i$, with the limit 
$r_i\gamma_i\rightarrow 0$, can also eliminate the removable 
singularity.    

There are two contributions to $a_{n,m}$ \cite{Chalmers1}.  The first 
one is, 

\bqr 
 \lambda \sum_{j=3}^\infty \sum_{a=1}^{j-1}(-1)^a 
   {1\over a!(j-a)!}  {a!\over \alpha_2!(a-\alpha_2)!}
\fqr 
\bqr  
\times 
\sum_{\alpha_1,\alpha_2;\beta_i}^{n=\alpha_1+\sum \beta_i}~
~\sum_{\tilde p=m+a-\alpha_2; p_i} ~ 
 (-1)^{{\tilde p}+\alpha_2}
  \prod_{i=1}^{\alpha_2} {{p_i^{p_i-2}}\over (p_i-1)!} 
\fqr 
\bqr 
{(\alpha_2+j-2a)!\over (\alpha_2+j-2a-\alpha_1)!} 
 \delta_{\alpha_2+j-2a,\alpha_1} 
 \times \prod_{i=1}^{\alpha_2} 
 2^{-p_i} \sum_{q=0}^{p_i} {p_i!\over q!(p_i-q)!} (-1)^{p_i-q} 
\fqr 
\bqr 
  \sum_{r=0}^m {m!\over (m-r)!} 
~\sum_{\gamma_i}  
  \prod_{i=1}^{\alpha_2} {\beta_i!\over (\beta_i-\gamma_i)!} 
 q^{\beta_i-\gamma_i} (-p)^{\gamma_i} 
 \vert_{\sum \gamma_i=b-a+{\tilde p}-r} \ .  
\label{counting}
\fqr 
The second one is found from setting the numbers to $j=0$, $a=0$, $b=2$ and 
$j=1$, $a=0$, $b=1$ and $j=2$, $a=0$, $b=0$ in the formulae of (21)-(25) 
of \cite{Chalmers1}.  This results in \rf{counting} by changing $\alpha_2 
\rightarrow \alpha_2+b$ and $\tilde p\rightarrow \tilde p+b$ with the 
values of $a$.  

These expansions are generally not Taylor, as the infinite number of 
derivatives about the origin of the K\"ahler potential might not converge.  
There are expansions of $x^\delta$ involving Laguerre polynomials which 
are polynomials in $x$; these derivatives of $x$ for small $\delta$ do 
not exist except by analytic continuation.  This is an example of the 
multiple sums might converge to radicals in the lowest order terms of 
the $\phi_i,\bar\phi_i$ expansion. 

The sums required to describe the K\"ahler potential might possess a 
hidden symmetry leading at a simplified form.  Mirror symmetry could 
indicate this.  However, even for low dimensional examples of torics 
such as simple Hirzebruch surfaces the metric forms are somewhat 
complicated.  

The full form of the expansion of the K\"ahler expansion about zero 
is akin to solving for the roots of a coupled set of polynomials.  It 
is interesting that the solution can be expressed in closed form, as there are 
theorems stating under certain functional forms that this is not possible.  
Performing the sums in \rf{counting} is very relevant; modular functions 
at specific values could play a role in summing these partitions also with 
values of $n$ and $m$. 

\vskip .2in 
\noindent{\it Discussion}  

A generating function for the toric K\"ahler potentials is given, pertaining 
to both finite and infinite dimensional quotients.  
The explicit form of the potential, and metrics, is produced in terms 
of sums over rational numbers.  This derivation is equivalent to solving 
the D-term constraints for a generic ${\cal N}=2$ gauged non-linear sigma 
model, which involves solving a system of specific coupled algebraic 
equations following from the charges and gaugings of the model.  Instead 
of transcendentally solving these equations, the counting of classical 
scalar field diagrams enables the computations of the solutions.

On the mathematical side, this derivation is equivalent to solving some 
complicated systems of algebraic equations.  A closed form is given to 
these solutions.  It appears that a more modular construction would 
simplify the results.  

The explicit form of the K\"ahler potentials should reveal more structure 
in the duality of the string models, and quantum field models, that give 
rise to these non-linear sigma models.  The specific form of the K\"ahler 
potential has clear applications to particle physics models.     

The explicit form of the toric metrics also leads to further information 
in the construction of solutions to generic algebraic systems 
\cite{Chalmers9}-\cite{Chalmers13}.  Possible simplifications in the 
handling of transcendental informaton are evident.  

In addition, the finding of integer solutions to polynomial solutions 
is relevant to the solutions of statistical mechanics problems 
\cite{Chalmers10}. 

\vskip .2in 
{\it Note added}: The metric form may be found at, and around, integer 
solutions $\phi_i=n_i$ by shifting of $\phi_i={\tilde \phi}_i+n_i$ 
coordinates 
and redoing the classical graph count.  If these integer points exist in 
the manifold, then there should be a locally flat space representation, 
which also results in the appropriate monodromy in the K\"ahler potential, 
as at $\phi_i=0$.  Counting the 
integer points appears possible due to the K\"ahler 
potential's expansion about these points.  The explicit this count, of the 
algebraic system modeled by the toric, requires some simplifications of 
the summations in order to find the appropriate flat space limit.

\vfill\break

\end{document}